
\magnification=1200
\font\cmbxxii=cmbx12

\def\i{\int_0^1}
\def\pl{\partial_\parallel}
\def\pt{\partial_\perp}
\def\D{\Delta_0}
\def\e{\varepsilon}
\def\f{\phi}
\def\lrl{\log r\Lambda}
\def\lal{\log a\Lambda}
\def\k{{\bf k}}
\def\X{{\bf X}}

{\nopagenumbers
\line{\hfil HUB-IEP-94/34}
\line{\hfil hep-th/9412123}
\vskip 1cm
\centerline{\bf GEOMETRICAL PROPERTIES OF CONFORMAL FIELD THEORIES}
\vskip .2cm
\centerline{\bf COUPLED TO TWO-DIMENSIONAL QUANTUM GRAVITY}
\vskip 1.5cm
\centerline{{\bf S.
Braune}{\footnote{*}{elm: braune@ifh.de}} }
\vskip .5cm
\centerline{\it Humboldt-Universit\"at Berlin}
\centerline{\it Mathematisch-Naturwissenschaftliche Fakult\"at 1}
\centerline{\it Institut f\"ur Physik, Invalidenstra\ss e 110}
\centerline{\it 10099 Berlin, Germany}
\vskip 1.5cm
\centerline{December, 1994}
\vskip 4cm
\centerline{\bf Abstract}
\vskip 1cm
In this work we discuss an approach due to F. David to the geometry
of world sheets of non-critical strings in quasiclassical approximation.
The gravitational dressed conformal dimension is related to the scaling
behavior of the two-point function with respect to a distance variable.
We show how this approach reproduces the standard gravitational
dressing in the next order of perturbation theory.
With the same technique we calculate the intrinsic Hausdorff dimension
of a world sheet.

\vfill\eject}

\centerline{\cmbxxii 1. Introduction}
\vskip .3in

In the theory of non-critical strings the change of some physical parameters
which is caused by quantum gravity plays an important role. This change is
called  gravitational dressing. In this paper we calculate the gravitational
dressing of scaling dimensions of Green functions with respect to their
distance behavior.
In chapter 2 we will briefly recall how to calculate $N$-point functions in a
conformal two-dimensional theory coupled to quantum gravity. We will see, that
studying Green functions with respect to distance has no physical sense
because after integration over all metrics the distance variable has no
longer a physical meaning. Technically, it can be seen as follows. To obtain
finite expressions for the Green functions which are independent of a
two-dimensional reference metric we have to renormalize our operators in such
a way, that the renormalized operators become operators of conformal
dimension  $(1,1)$. Then the Green functions have the distance behavior of
$(1,1)$ operators, which has nothing to do with the bare conformal dimension.
Only the exponent of the Liouville mass depends on the bare conformal
dimension [1,2,3,4,5]. This exponent is usually called dressed scaling
dimension. The relation between the bare and dressed dimensions is given by
the Knizhnik-Polyakov-Zamolodchikov- (KPZ)-formula [6].
In chapter 3 we shortly explain the idea of F. David [7],
how to define a two-point function depending on the geodesic distance.
It is done by restriction of the functional integral over the metric field in
such a way, that the geodesic distance between two points is fixed for
all metrics over which is integrated. This fixed geodesic distance becomes the
distance variable of the two point function. We will see, that this Green
function has a power-like distance behavior with an exponent which is similar
to the dressed dimension of the operators. In the quasiclassical limit this
exponent is in lowest order of perturbation  theory in agreement with the
KPZ-relation [7].
But there are still some open questions in this approach. It becomes not
clear in detail, how this Green function is defined exactly. Furthermore, we
have to take into account that the geometrical objects like the geodesic line
and the geodesic distance are composite operators which have to be
renormalized. Of course we expect that the renormalization becomes more
complicated if we calculate expectation values not only of vertex operators.
So in principle, we have to find a renormalization which removes all
divergences coming from the self-contractions at the vertices and from the
other composite operators simultaneously. Leaving aside the question of
renormalization we try to find Green functions which reproduce the scaling
in their distance behavior with exponents given by the KPZ-relation.
We use the classical formula for the geodesic distance and find two possible
candidates for geometrical two-point functions in this way.
A decision which Green function is the adequate one can be made in the next
order of perturbation theory only. We do this in the chapters 4-6.
Furthermore, it is an open question, whether the free field approximation can
be used. This question we do not address in this paper.

In chapter 5 we calculate the intrinsic Hausdorff dimension of the string
world sheet applying the same perturbation theory. The Hausdorff dimension is
an useful characteristic of a sheet or surface. The calculation is done
because it is expected that for $D\to +1$ the world sheet gets deformations
like spikes.

\vfill\eject
\centerline{\cmbxxii 2. Usual definition of gravitational dressed conformal
dimension}
\vskip .3in

The usual way for calculating N-point functions of vertex operators in
noncritical string theory was worked out in [1,2,3,4]. One has to calculate
$$ \Bigl\langle\prod_j \Psi_{\k_j}(z_j)\Bigr\rangle=
\int Dg_{ab}D_g\X\Bigl(\prod_j e^{i\k_j \X(z_j)}\Bigr)
e^{-{1\over{8\pi}}\int d^2 x
\sqrt{g} g^{ab}\partial_a \X\partial_b \X} .\eqno(2.1)$$
$\X$ is the string position, depending on the 2D-coordinate $z$. The
metric on the world sheet is $g_{ab}$. We parametrize it by taking $g_{ab}=
e^\phi\hat{g}_{ab} $, where $\hat{g}_{ab}$ is a reference metric.
$\Psi_{\k_j}(z_j) = e^{i\k_j \X(z_j)}$ is called vertex operator.

Now we restrict ourselves to the topology of the sphere,
$\hat g_{ab}=\delta_{ab}$. The propagator of a free field
in this conformal flat background metric $g_{ab}=e^\phi \delta_{ab}$ is
$ \langle \X(z_1)\X(z_2)\rangle_{\!\f} =
-\log\vert z_2 - z_1 \vert^2
- {1\over 2}\f(z_1) - {1\over 2}\f(z_2) + {\rm const} $.
To calculate Green functions we perform the functional integration over the
free string position $\X$ and over the diffeomorphism ghosts and make the
integration measure translationally invariant. After rescaling
$\sqrt{{25-D}\over 3}\phi=2\varphi$ we obtain for the partition function
of the Liouville sector
$$Z = \int D_{\hat{g}}\varphi \;e^{-{1\over{8\pi}}\int d^2 x\sqrt{\hat{g}}
\bigl(\hat{g}^{ab}\partial_a\varphi\partial_b\varphi+\hat{R}Q\varphi
+\mu^2e^{\alpha\varphi}\bigr)} \eqno(2.2)$$
with $Q^2={{25-D}\over 3}$. To ensure the independence of $Z$ from the
reference metric $\hat{g}$ one has to set $\alpha = {{Q-\sqrt{Q^2-8}}
\over 2}$. To get independence of $\hat{g}$ also for correlation functions
one has to replace $\Psi_\k(z)$ by
$$[\Psi_\k(z)] = e^{\beta(\k) \varphi(z)}e^{i\k\X(z)}.\eqno(2.3)$$
Then one gets an equation for $\beta(\k)$:\
${1\over 2}\beta(\k)\bigl(Q-\beta(\k)\bigr)+{\k^2\over 2}=1$ or
$\beta = {Q-\sqrt{Q^2+8({\k^2\over 2}-1)}\over 2}$.
Without quantizing $g_{ab}$ the conformal dimension was
$\D={\k^2\over 2}.$ We do not obtain any restriction for the
Liouville mass $\mu$. For the N-point function one finds
$$\Bigl\langle\prod_j[\Psi_{\k_j}(z_j)] \Bigr\rangle \sim
(\mu^2)^{-{1\over\alpha}\sum_j\beta (\k_j)}. \eqno(2.4)$$
By construction the dependence on $\vert z_j - z_i\vert$
is the trivial one as usual in conformal field theory of operators with
conformal dimension 1. The exponent of the Liouville mass $\mu$ is related to
the gravitational dressed scaling dimension of the corresponding vertex
operator
$\Delta = 1 - {\beta (\k)\over\alpha}.$ We obtain the well known KPZ-relation
$$\Delta = {{\sqrt{Q^2+8(\D-1)}-\sqrt{Q^2-8}}\over
{Q-\sqrt{Q^2-8}}}. \eqno(2.5)$$
Another way to get this result based on the investigation of the
renormalization properties is shown in [8,9].

We can make an expansion around the quasiclassical limit $D
\rightarrow -\infty$, that is equivalent to $Q\rightarrow\infty$.
Then the KPZ-relation becomes
$$\Delta =\D-{{2\D(\D-1})\over {Q^2}}+{{4\D(\D-1)(2\D-3)}\over{Q^4}}+
O({1\over{Q^6}}).\eqno(2.6)$$

\vskip .3in
\centerline{\cmbxxii 3. The geometrical approach}
\vskip .3in

We have to remember that the gravitational dressed conformal dimension has
nothing to do with the power-like behavior of the N-point function with
respect to the distance of the points, because there is no possibility to
define this distance after integrating over the metric field. That is why F.
David proposed another 2-point function by functional integrating only over
all metric fields yielding a fixed geodesic distance $r$ between the points
$x$ and $y$ [7]
$$G(r)=\biggl\langle
\int d^2 x\int d^2 y\sqrt{g(x)}\sqrt{g(y)} \;
\delta(r-d_\f(x,y))\Psi(x)\Psi(y)\biggr\rangle.\eqno(3.1)$$
We substitute the $\theta-$ function in [7] by a $\delta-$ function.
This changes the power-like behavior by $-1$ in the exponent.
Furthermore we drop the normalization denominator.
This two-point function has a natural relation to the geodesic distance.
$\Psi$ denotes the vertex operator $e^{i\k\X}$. The functional integration
over $\X$ has to be performed carefully because of the divergences coming from
the self contractions at the vertices. If we remove them as a whole we get
$$\langle \Psi(x)\Psi(y)\rangle_{\!\f}=e^{-\D
(\phi(x)+\phi(y))}\vert y- x\vert^{-4\D}.\eqno(3.2)$$
If we remove the divergent part only we get
$$\langle \Psi(x)\Psi(y)\rangle_{\!\f}=\vert y- x\vert^{-4\D}.\eqno(3.3)$$
In both cases we make use of the momentum conservation.
We will use the first version in chapter 5 and the second version in chapter
6. We still have to do the integration over the Liouville field.

At first we need an expression for $\delta(r-d_\f(x,y))$.
Later we will see that an expansion around
the quasiclassical limit is equivalent to the power series in $\phi$.
For calculating the geodesic distance between two points $x$ and $y$ in the
metric $g_{ab}=e^{\phi}\delta_{ab}$ we have to solve the well-known geodesic
equation $-\ddot\xi^a(v)=\Gamma_{bc}^a(\xi(v))\dot\xi^b(v)
\dot\xi^c(v)$ with $\xi(0)=x$ and $\xi(1)=y$, which becomes
$$-\ddot\xi^a (v)=\partial_b\phi(\xi(v))\;\dot\xi^b (v)
\dot\xi^a (v)-{1\over 2}\partial_a\phi(\xi(v))\;\dot\xi^b (v)
\dot\xi^b (v)\eqno(3.4)$$ in this conformal flat metric.
It can be solved by iteration $$\xi_{(p+1)}^a(v)=\xi_{(0)}^a+\i du\;G(v,u)
\Gamma_{bc}^a(\xi_{(p)}(u)) \dot\xi_{(p)}^b (u) \dot\xi_{(p)}^c (u).
\eqno(3.5)$$
$G$ is the Green function for the second derivative $G(v,u)=\min(v,u)-vu$,
and we take $\vec \xi_{(0)}(v)=\vec x+v(\vec y-\vec x)$.
We use the coordinates $\xi^\parallel=
{{(\vec y-\vec x)\vec\xi}\over{\vert \vec y-\vec x\vert}}$,
$\xi^\perp={{(\vec y-\vec x)\times\vec\xi}\over
{\vert \vec y-\vec x\vert}}$ and
$\pl\f={{(\vec y-\vec x)\nabla\f}\over{\vert \vec y-\vec x\vert}}$,
$\pt\f={{(\vec y-\vec x)\times\nabla\f}\over{\vert \vec y-\vec x\vert}}.$
We denote the euclidean distance in the coordinate space by
$d=\vert \vec y-\vec x\vert$.

Now we need the Green functions of the Liouville field. In the paper of F.
David it was shown how to obtain an effective action for the Liouville field.
We replace the Liouville mass $\mu$ via a Laplace transformation by an
area variable $A$
$$Z=\int_0^\infty dA \;e^{-{\mu^2\over{8\pi}}A}Z(A)\eqno(3.6)$$ with
$$Z(A)=\int D\varphi \;e^{-{1\over{8\pi}}\int d^2 x \sqrt{\hat{g}}
(\hat{g}^{ab}\partial_a\varphi\partial_b\varphi+\hat{R}Q\varphi)}
\delta\Bigl(A-\int d^2x\sqrt{\hat{g}}e^{\alpha\varphi}\Bigr).\eqno(3.7)$$
Note that $Z(A)$ has to be calculated with $\mu=0$. Performing the
integration over the zero mode and writing the remaining integrand as
an exponential one gets the following
effective action for the Liouville field
$$\eqalignno{S_{\rm eff}={1\over{8\pi}}\int d^2 x \sqrt{\hat{g}}
\Biggl(&\hat{g}^{ab}\partial_a\tilde{\varphi}\partial_b\tilde{\varphi}+
\hat{R}Q\tilde{\varphi}- &(3.8)\cr
-&{{4\pi(1-h)Q\alpha^2}\over{\hat{A}}}\;\tilde{\varphi}^2-
{{4\pi(1-h)Q\alpha^3}\over{3\hat{A}}}\;\tilde{\varphi}^3-
{{\pi(1-h)Q\alpha^4}\over{3\hat{A}}}\;\tilde{\varphi}^4\Biggr)+ \cr
+{1\over{8\pi}}\int d^2 x\sqrt{\hat{g}}&\int d^2 y\sqrt{\hat{g}}\;
{{\pi(1-h)Q\alpha^4}\over{\hat{A}^2}}\;\tilde{\varphi}^2(x)
\tilde{\varphi}^2(y)+O(\tilde{\varphi}^5).}$$
In the limit $\hat{A}=\int d^2x\sqrt{\hat{g}}\rightarrow\infty$ we can take
$\tilde{\varphi}$ as an massless free field.
$\Lambda=\hat{A}^{-{1\over 2}}$ acts as an IR regulator
$$\eqalignno{\bigl\langle\f(z_1)\f(z_2)\bigr\rangle =
-{8\over Q^2}&\log\vert z_2- z_1\vert\Lambda, &(3.9)\cr
\bigl\langle\f(z_1)\f(z_2)\f(z_3)\f(z_4)\bigr\rangle =
{64\over Q^4}&\bigl(\log(\vert z_2- z_1\vert\Lambda)
\log(\vert z_4- z_3\vert\Lambda)+\cr
&+\log(\vert z_3- z_2\vert\Lambda)\log(\vert z_4- z_1\vert\Lambda)+\cr
&+\log(\vert z_3- z_1\vert\Lambda)\log(\vert z_4- z_2\vert\Lambda)\bigr).}$$
We see that the expansion in powers of $\f$ is equivalent to the expansion
around the quasiclassical limit in powers of ${1\over Q}$.

The calculation of the Green function was done up to the order
${1\over Q^2}$ with the result $G(r)\sim r^{-4\Delta}$ with $\Delta$
obeying the KPZ-relation.

In addition the volume of a disk in the string world sheet
$$\langle V(r)\rangle =
\biggl\langle\int d^2 x\sqrt{g(x)}\theta(r-d_\f(x,o))\biggr\rangle
\eqno(3.10)$$
was calculated in [7] up to the order ${1\over Q^2}$ with the result
$\langle V(r)\rangle\sim r^2$.
Similar with the help of the Gauss-Bonnet theorem the number
of connected components of a circle was calculated with the result
$\langle n(r)\rangle\sim r^0$. There were not found any spikes in the world
sheet by expanding around the quasiclassical limit.

In our opinion there are still some open questions in this context.
At first it has to be checked whether the  KPZ-relation
is fulfilled by $\Delta$ defined via the $r$ scaling of $G(r)$
in higher orders. We know that in order ${1\over Q^2}$ the factor
$\sqrt{g} e^{-\D\f}$ coming from the area element and from $(3.2)$ can be
replaced by the dressing factor $e^{\beta\varphi}$ without changing the
exponent in $G(r)\sim r^\Delta$ up to the calculated order. The formulas
will be shown in chapter 6.
So we are not sure which definition for the distance-dependent two-point
function leeds to the scaling behavior of the standard dressing [1,2,3].
These question can be answered only by continuing
the perturbation theory to the next nontrivial order.

Furthermore, we would like to mention that there is a possibility to calculate
the Green function of the Liouville field exactly [10,11,12,13].
The two- and three point functions of exponentials of the
Liouville field are known. From these in principle one can calculate
the Green functions for the Liouville field itself by differentiation.
However, there are still unsolved problems with an implied limiting process.
If we would take the exact Green function, it would be enough for the
calculation of the next order corrections to do this up to the order ${\f^3}$.
However, we try to do it use the free field propagators. Therefore we have
to calculate it up to $\f^4$.

\vskip .3in
\centerline{\cmbxxii 4. Next steps in calculating higher orders}
\vskip .3in

The first thing we need for continuing this calculation is a power series
of the function $F_r(d_\f(x,y))=\theta(r-d_\f(x,y))$ in powers of
$\phi$. We need the geodesic distance between the points $x$ and $y$. To this
purpose we expand around the line $\xi_{(0)}(v)=x+v(y-x)$.
We set $d=\vert y-x \vert$ and write $\f_s$ instead of
$\f(\xi_{(0)}(s))$ and $G_{t,s}$ instead of $G(t,s)$.
The length of this curve is $d_\f=\i ds\sqrt{g_{ab}\dot\xi^a\dot\xi^b}$.
Our result up to the order $\f^4$ is
$$\eqalignno{& &(4.1)\cr
d_\f &(x,y)=d\i dv\biggl(1+{\f_v\over 2}+{\f_v^2\over 8}+
{\f_v^3\over 48}+{\f_v^4\over 384}\biggr)
-{d^3\over 8}\i dv\i du \;G_{v,u}\pt\f_u\pt\f_v -\cr
&-{d^3\over 8}\i dv\i du \;G_{v,u}\pt\f_u\f_v\pt\f_v
+{d^3\over 16}\i dv\biggl(\i du\; G_{v,u\vert v}
\pt\f_u\biggr)^2\f_v +\cr
&+{d^5\over 16}\i dv\biggl(\i du\;G_{v,u}\pt\f_u\biggr)^2
\pt\pt\f_v-\cr
&-{d^3\over 32}\i dv\i du\;G_{v,u}\pt\f_u\f_v^2\pt\f_v
-{d^3\over 32}\i dv\i du\;G_{v,u}\f_u\pt\f_u
\f_v\pt\f_v+\cr
&+{d^3\over 32}\Biggl(\i dv\i du\; G_{v,u\vert v}
\pt\f_u\f_v\Biggr)^2+\cr
&+{d^3\over 16}\i dv\biggl(\i du\; G_{v,u\vert v}
\pt\f_u\biggr)\biggl(\i dt\; G_{v,t\vert v}
\f_t\pt\f_t\biggr)\f_v-\cr
&-{d^3\over 64}\i dv\biggl(\i du\; G_{v,u\vert v}
\pt\f_u\biggr)^2\f^2_v-\cr
&-{d^5\over 16}\i dv\biggl(\i du\;
G_{v,u\vert v}\pt\f_u\biggr)
\biggl(\i dt\i ds\; G_{v,t\vert v}G_{t,s}
\pt\f_s\pt\pt\f_t\biggr)\f_v-\cr
&-{7d^5\over 128}\i dv
\biggl(\i du\; G_{v,u\vert v}\pt\f_u\biggr)^2
\biggl(\i dt\;G_{v,t}\pt\f_t\biggr)\pt\f_v+\cr
&+{d^5\over 16}\i dv\biggl(\i du\;G_{v,u}\pt\f_u\biggr)
\biggl(\i dt\;G_{v,t}\f_t\pt\f_t\biggr)\pt\pt\f_v+\cr
&+{d^5\over 32}\i dv\biggl(\i du\;G_{v,u}\pt\f_u\pt\f_v\biggr)^2
+{d^5\over 32}\i dv\biggl(\i du\;G_{v,u}\;
\pt\f_u\biggr)^2\f_v\pt\pt\f_v-\cr
&-{d^7\over 32}\i dv\i du\i dt\i ds\;G_{v,u}G_{u,t}G_{t,s}
\pt\f_s\pt\pt\f_t\pt\pt\f_u\pt\f_v-\cr
&-{d^7\over 96}\i dv\biggl(\i du\;G_{v,u}\pt\f_u\biggr)^3
\pt\pt\pt\f_v+O(\phi^5).}$$

Then we need a prescription how to regularize the propagator.
In [7] was used a combination of cut-off and finite-part regularization.
Logarithmical divergences should be regulatized by a cut-off
$\e={a\over d}$. Linear and higher divergences should be regularized by
the prescription $\int_0^x {dy\over y^\alpha}={x^{\alpha+1}\over{\alpha+1}}$
for every $\alpha\geq 2$. We have to notice that this hybrid regularization
is not unique. The result depends on the order of the integrals.
We see this if we regularize the expectation value of the classical identity
$$\eqalignno{2\biggl(\i ds\;\f_s\biggr)\biggl(\i dv\i du\i dt&\;G_{v,u\vert v}
G_{u,t\vert u}\pt\f_t\pt\f_u\f_v\biggr)= &(4.2)\cr
=&\biggl(\i ds\;\f_s\biggr)\biggl(\i dv\Bigl(\i du\; G_{v,u\vert v}
\pt\f_u\Bigr)^2\f_v\biggr)- \cr
&-\biggl(\i ds\;\f_s\biggr)^2
\biggl(\i dv\i du \;G_{v,u}\pt\f_u\pt\f_v\biggr).}$$
However, a smooth regularization
$\bigl\langle\f(s\vec x)\f(t\vec x)\bigr\rangle \to
-{8\over Q^2}\Bigl(\log\vert\vec x\vert\Lambda
+{1\over 2}\log\bigl((t-s)^2+\e^2\bigr)\Bigr)$
which would be unique becomes too complicated for analytic integration.
We make the regularization unique by a substitution of the propagator by
$$\bigl\langle\f(x)\f(y)\bigr\rangle=
-{8\;\theta(a-\vert y- x\vert)\over Q^2}\log a\Lambda
-{8\;\theta(\vert y- x\vert-a)\over Q^2}
\log{\vert y- x\vert\Lambda}\eqno(4.3)$$
\vfill\eject
and obtain the following derivatives:
$$\eqalignno{&\bigl\langle\pt\f(x)\f(y)\bigl\rangle=0,&(4.4) \cr
&\bigl\langle\pt\f(x)\pt\f(y)\bigr\rangle=
{8\;\theta(\vert y- x\vert-a)\over {Q^2\vert y- x\vert^2}},
\;\;\;\;\;
\bigl\langle\pt\pt\f(x)\f(y)\bigr\rangle=
-{8\;\theta(\vert y- x\vert-a)\over {Q^2\vert y- x\vert^2}},\cr
&\bigl\langle\pt\pt\f(x)\pt\f(y)\bigl\rangle=0, \cr
&\bigl\langle\pt\pt\f(x)\pt\pt\f(y)\bigr\rangle=
{48\;\theta(\vert y- x\vert-a)\over {Q^2\vert y- x\vert^4}}-
{24\;\delta(\vert y- x\vert-a)\over {Q^2 a^3}},\cr
&\bigl\langle\pt\pt\pt\f(x)\pt\f(y)\bigr\rangle=
-{48\;\theta(\vert y- x\vert-a)\over {Q^2\vert y- x\vert^4}}+
{24\;\delta(\vert y- x\vert-a)\over {Q^2 a^3}}.}$$
Then the regularization of $(4.2)$ is unique. After integrating we ignore
all power-like divergences, we are only looking for finite terms and
logarithmical divergences. We have to notice that we do not remove any self
contractions of the Liouville field.

Alternatively there is another method for the calculation of the
geodesic distance. It is based on a replica field [7]. The
cut-off-regularization in this combined system of Liouville and
replica field does not have any linear divergences.
We are afraid, that the Feynman integrals of the next
order become too complicated for the analytic calculation.

\vskip .3in
\centerline{\cmbxxii 5. Calculation of the intrinsic Hausdorff dimension}
\centerline{\cmbxxii and of a Green functions}
\vskip .3in

We want to calculate the power-like behavior of $\langle V(r)\rangle\sim
r^{\Delta_H}$. Our result is that all logarithmical divergences cancel.
After dropping all terms $\;\sim\e$ and $\;\sim{1\over\e}$ we can factorize
our result in $r^2$ and a factor which is independent of $r$. That means
$\Delta_H=2+O({1\over Q^6})$. This is in agreement with the theorem, that the
Hausdorff dimension is never smaller than the topological dimension.
The extrinsic Hausdorff dimension of the same world sheet was calculated
in [2]. The relation between intrinsic and extrinsic Hausdorff dimension
[14] is not valid because in the quasiclassical limit there is not an
embedding space.

Now we calculate the Green function $(3.1)$ proposed by F. David without
using the dressing coefficients of chapter 2 and the two-point function
$(3.2)$. The two-point function which gives the
KPZ-formula at least in the lowest order is
$$G(r)=\Biggl\langle\int d^2 x\int d^2 y\;{{\sqrt{g(x)g(y)}\;
\delta(r-d_\f(x,y))}\over{e^{\D(\f(x)+\f(y))}\;
\vert y- x\vert^{4\D}}}\Biggr\rangle. \eqno(5.1) $$
Our result for $G(r)$ can be factorized
$$\eqalignno{
G(r)=-2\pi & r^{1-4\D}\hat{A}\biggl( 1-{4-16\D +8\D^2\over Q^2}
-{4-8\D(1-\D)\over Q^2}\lal\biggr)\cdot &(5.2) \cr
&\cdot\biggl(1-{8\D(1-\D)\over Q^2}\lrl-\cr
&-{16\D(1-\D)\over Q^4}\lrl
+{(8\D(1-\D))^2\over 2Q^4}(\lrl)^2\biggr).}$$
With $e^{\Delta\lrl}=(r\Lambda)^\Delta$ we get for the exponent of the
Green function
$$G(r)\sim r^{1-4(\D+{2\D(1-\D)\over Q^2}
+{4\D(1-\D)\over Q^4})}. \eqno(5.3) $$
This exponent disagrees with the KPZ-formula.
So we conclude that it is not possible to obtain a distance dependent
Green function with this scaling behavior in this way.

\vskip .3in
\centerline{\cmbxxii 6. Two-point functions including the dressing factor}
\vskip .3in

Now we introduce Green functions, which explicitly include the dressing
factor $e^{\beta\varphi}$ explained in chapter 2. Now we have to use the
two-point function $(3.3)$, if we want to obtain the scaling of standard
dressing.  We have to calculate the function
$G_d(r)=\bigl\langle\int d^2 x\int d^2 y\;
[\Psi(x)][\Psi(y)]\;\delta(r-d_\f(x,y))\bigr\rangle$, which gives
$$G_d(r)=\Biggl\langle\int d^2 x\int d^2 y\;{e^{{Q\beta\over 2}(\f(x)+\f(y))}\;
\delta(r-d_\f(x,y))\over{\vert y- x\vert^{4\D}}}\Biggr\rangle. \eqno(6.1) $$
In the quasiclassical limit the dressing coefficient is
${Q\beta\over 2}=(1-\D)+2{(1-\D)^2\over Q^2}+8{(1-\D)^3\over Q^4}+O(Q^{-6})$.
Using the same perturbation theory and regularization as before we get
$$\eqalignno{
G_d (r)=-2\pi & r^{1-4\D}\hat{A}\biggl( 1-{4-16\D +8\D^2\over Q^2}
-{4-8\D(1-\D)\over Q^2}\lal\biggr)\cdot  &(6.2)\cr
&\cdot\biggl(1-{8\D(1-\D)\over Q^2}\lrl-\cr
&-{16\D(3-5\D+2\D^2)\over Q^4}\lrl
+{(8\D(1-\D))^2\over 2Q^4}(\lrl)^2\biggr).}$$
The factor with the $\lrl-$terms does not depend on the UV-cut-off $a$.
The $r-$dependent part of $G_d$ can be written as an exponential
$$ G_d(r)\sim r^{1-4(\D+{2\D(1-\D)\over Q^2}
+{4\D(1-\D)(3-2\D)\over Q^4})}.\eqno(6.3)$$
We see that this Green function $G_d(r)$ perturbatively gives an exponent
obeying the KPZ-relation. Now $\Delta$ is the exponent in the power-like
dependence on the distance.

In this calculation we did not pay attention to the conservation of the
charge. If we did this we would have $G_d(r)=\Bigl\langle\int d^2 x\int d^2 y\;
e^{{Q\over 2}((Q-\beta)\f(x)+\beta\f(y))}\;\delta(r-d_\f(x,y))
\Psi(x)\Psi(y)\Bigr\rangle $. Expanding the exponentials and calculating
the expectation values we would obtain the following formulas
$\bigl\langle e^{{Q^2\over 2}\f}\bigr\rangle=e^{-Q^2\lal}$,
$\bigl\langle e^{{Q^2\over 2}\f}\f_1\bigr\rangle=
{Q^2\over 2}\langle\f\f_1\rangle e^{-Q^2\lal}$, and
$\bigl\langle e^{{Q^2\over 2}\f}\f_1\f_2\bigr\rangle=
\bigl(({Q^2\over 2})^2\langle\f\f_1\rangle\langle\f\f_2\rangle+
\langle\f_1\f_2\rangle\bigr)e^{-Q^2\lal}$.
Using $\bigl\langle\f(z_1)\f(z_2)\bigr\rangle =
-{8\over Q^2}\log{{\vert z_2- z_1\vert}\Lambda}$ we see that
the power series in $\f$ do not causes a power Series in ${1\over Q}$.
So our method can not be applied to $(6.4)$. However, in the complete
theory with $\mu\ne 0$ this charge conservation is not valid.

\vskip .3in
\centerline{\cmbxxii 7. Conclusions}
\vskip .3in

The Green function $G_d(r)$ defined in $(6.1)$ gives a power series in
$1\over Q^2$ which can be written as $r^{1-4\Delta}$, if one uses the
effective free field action for the Liouville field and if one removes
the linear and higher divergences. The logarithmic divergent factor in
$(6.2)$ and the power-like divergences which we dropped have to be
investigated carefully. There should be an additional renormalization
factor for the composite fields like the geodesic distance given in
$(4.1)$. This renormalization should include a unique prescription
for removing self contractions. It should remove all divergences without
destroying the power-like distance behavior.

\vskip .3in
\centerline{\cmbxxii Acknowledgements}
\vskip .3in

I would like to thank H. Dorn for many helpful discussions
and D. L\"ust for his support.

\vskip .3in
\centerline{\cmbxxii References}
\vskip .3in

\item{[1]} J. Distler, H. Kawai, {\it Nucl. Phys.} {\bf B 321} (1989) 509,
\item{[2]} J. Distler, Z. Hlousek, H. Kawai, {\it Int. J. Mod. Phys.}
{\bf A 5} (1990) 1093,
\item{[3]} N. Seiberg, {\it Progr. Theor. Phys. Suppl.} {\bf 102} (1990) 319,
\item{[4]} F. David, {\it Mod. Phys. Lett.} {\bf A 3} (1988) 1651,
\item{[5]} E. Abdalla, M. C. B. Abdalla. D. Dalmazi, A. Zadra, {\it
2D-Gravity in Non-Critical Strings}, Lecture
Notes in Physics {\bf m 20}, Springer-Verlag Berlin, Heidelberg (1994),
\item{[6]} V. G. Knizhnik, A. M. Polyakov, A. B. Zamolodchikov,
{\it Mod. Phys. Lett.} {\bf A 3} (1988) 819,
\item{[7]} F. David, {\it Nucl. Phys.} {\bf B 368} (1992) 672,
\item{[8]} H. Dorn, H.-J. Otto, {\it Phys. Lett.} {\bf B 280} (1992) 204,
\item{[9]} H. Dorn, H.-J. Otto, {\it Phys. Lett.} {\bf B 232} (1989) 327,
\item{[10]} H. Dorn, H.-J. Otto, {\it Nucl. Phys.} {\bf B 429} (1994) 375,
\item{[11]} Vl. S. Dotsenko, V. A. Fateev, {\it Nucl. Phys.} {\bf B 240}
(1984) 312,
\item{[12]} Vl. S. Dotsenko, {\it Mod. Phys. Lett.} {\bf A 6} (1991) 3661,
\item{[13]} M. Goulian, M. Li, {\it Phys. Rev. Lett.} {\bf 66} (1991) 2051,
\item{[14]} T. Jonsson, {\it Phys. Lett.} {\bf B 278} (1992) 89.

\bye